\documentclass[aps,preprint]{revtex4}%
\usepackage{amssymb}
\usepackage{amsfonts}
\usepackage{amsmath}
\usepackage{graphicx}%
\providecommand{\U}[1]{\protect\rule{.1in}{.1in}}

\begin{document}
\title{Blackbody Radiation in Classical Physics: A Historical Perspective}
\author{Timothy H. Boyer}
\affiliation{Department of Physics, City College of the City University of New York, New
York, New York 10031}

\begin{abstract}
We point out that current textbooks of modern physics are a century
out-of-date in their treatment of blackbody radiation within classical
physics. \ Relativistic \textit{classical} electrodynamics including classical
electromagnetic zero-point radiation gives the Planck spectrum with zero-point
radiation as the blackbody radiation spectrum. \ In contrast, nonrelativistic
mechanics cannot support the idea of zero-point energy; therefore if
nonrelativistic classical statistical mechanics or nonrelativistic mechanical
scatterers are invoked for radiation equilibrium, one arrives at only the
low-frequency Rayleigh-Jeans part of the spectrum which involves no zero-point
energy, and does not include the high-frequency part of the spectrum involving
relativistically-invariant classical zero-point radiation. \ Here we first
discuss the correct understanding of blackbody radiation within relativistic
classical physics, and then we review the historical treatment. \ Finally, we
point out how the presence of Lorentz-invariant classical zero-point radiation
and the use of relativistic particle interactions transform the previous
historical arguments so as now to give the Planck spectrum including classical
zero-point radiation. Within relativistic classical electromagnetic theory,
Planck's constant $\hbar$ appears as the scale of source-free zero-point radiation.

\end{abstract}
\maketitle

\section{Introduction}

During the first two decades of the 20th century, the blackbody radiation
spectrum appeared prominently in physics research. \ Debates over the correct
theoretical interpretation of the experimental measurements reached a climax
around 1909. \ The view that emerged from the debates was that classical
physics led inevitably to the Rayleigh-Jeans spectrum and was incapable of
explaining the experimentally-measured blackbody spectrum. \ In this article,
we use historical accounts to go back to the debates of a century ago, but now
with the awareness of the developments within \textit{classical} physics of
the intervening years. \ We apply the new knowledge to the old controversies.
\ We point out why the old arguments fail; in some cases the arguments must be
abandoned, and in other cases the arguments are easily corrected in the light
of new information. \ A reanalysis of the conflict of the previous century
leads to the conclusion that classical physics can indeed give an accurate
account of the Planck spectrum appearing in the experimental data.

The textbooks of modern physics have not caught up with current knowledge
regarding blackbody radiation within classical physics and instead provide
students with an incorrect classical view.\cite{err} \ An accurate statement
of the situation is as follows: the use of relativistic physics with the
inclusion of classical electromagnetic zero-point radiation leads to the
Planck spectrum including classical zero-point radiation, whereas the use of
nonrelativistic classical statistical mechanics or of nonrelativistic
nonlinear scattering systems leads only to the low-frequency, Rayleigh-Jeans
part of the spectrum.\ \ The erroneous view that classical physics leads
inevitably to the Rayleigh-Jeans form for the entire spectrum corresponds to
the mistaken conclusion reached by the physicists of a century ago.\ However,
the physicists of the early 20th century were unaware of two crucial elements
of classical physics, namely 1) the presence of classical electromagnetic
zero-point radiation with a scale set by Planck's constant $\hbar,$ and 2) the
importance of special relativity.

Many of the readers of this article (having been trained using misleading
textbooks) may be thoroughly skeptical of the claims made in the preceding two
paragraphs. \ Therefore let me point out the obvious problem with the views
which are presented in contemporary treatments of blackbody radiation within
classical physics. \ Classical electrodynamics is a relativistic theory.
\ Nonrelativistic mechanics is not. \ However, electromagnetic radiation can
be brought to thermal equilibrium only by the interaction of electromagnetic
radiation with charged mechanical systems.\ \ How can a nonrelativistic
mechanical system which is inconsistent with special relativity be expected to
give the equilibrium spectrum of relativistic classical electrodynamics?
\ Thus classical statistical mechanics with its energy equipartition ideas is
a nonrelativistic theory which cannot support the concept of zero-point energy
for either mechanical particles or wave phenomena. \ In classical statistical
mechanics, a nonrelativistic particle of mass $m$ in one spatial dimension in
thermal equilibrium has a kinetic energy $KE=(1/2)mv^{2}=(1/2)k_{B}T.$
\ However, this means that for fixed temperature $T$ and sufficiently small
mass $m$, the velocity $v$ of the particle will exceed the speed of light in
vacuum $c.$ \ Such a situation is nonsense in a relativistic theory. \ The
application of nonrelativistic classical statistical mechanics or of
nonrelativistic scatterers to the problem of thermal radiation with the
expectation of deriving the full radiation spectrum is a fundamental error
which has persisted for over a century and still appears in the textbooks of
modern physics. \ Indeed, we are aware that thermal radiation involves two
regimes depending upon whether the ratio $\hbar\omega/(k_{B}T)$ is small or
large. \ When $\hbar\omega/(k_{B}T)$ is small, we are in the Rayleigh-Jeans
regime where nonrelativistic energy-equiparition ideas are satisfactory.
\ When $\hbar\omega/(k_{B}T)$ is large, we are in the region dominated by
Lorentz-invariant and scale-invariant classical zero-point radiation with a
scale set by Planck's constant $\hbar.$ \ Lorentz-invariant mechanics (which
depends on the constant $c)$ provides the appropriate transition between the
two regimes for scatterers of radiation, with the ratio $mc^{2}/(k_{B}T)$
providing the appropriate limits for a charged mass in a Coulomb potential.

The outline of our presentation is as follows. \ Before discussing the
historical situation, we first summarize our current understanding of some
basic classical ideas. \ We point out the experimental basis for classical
zero-point radiation. \ Then we discuss relativistic mechanical systems, and
we note the convenience of using action variables when discussing thermal
equilibrium. \ Next we turn to the historical aspects of blackbody radiation.
\ We summarize the thermodynamic aspects leading to the Wien displacement
theorem. \ We review the thermodynamics of a single radiation mode and note
that thermodynamics allows the asymptotic limits corresponding to zero-point
energy and energy equipartition. \ Next we describe the historical blackbody
controversies of the early 20th century, including Planck's search for a place
for his constant $\hbar$ within classical electrodynamics, and concluding with
the Rayleigh-Jeans consensus. \ In the last section, we review a number of
classical electromagnetic calculations of the early 20th century which led to
the Rayleigh-Jeans spectrum, and we show how the inclusion of classical
zero-point radiation and appropriate relativistic treatments transforms the
calculations into derivations of the Planck spectrum with zero-point radiation
within classical physics. \ A closing summary ends the discussion. \ 

\section{Classical Electromagnetic Zero-Point Radiation}

Before our discussion of the historical controversy involving blackbody
radiation within classical electromagnetic theory, we wish to emphasize the
current understanding of the two aspects which the early investigators missed:
zero-point radiation and the importance of relativity. \ We start with the
idea of classical electromagnetic zero-point radiation.

\subsection{Experimental Evidence for Classical Zero-Point Radiation}

The experimentalists who investigated the blackbody radiation spectrum around
the turn of the 20th century were able to measure only the thermal radiation
from their sources in excess of the ambient radiation surrounding their
detectors. \ If their sources were at the same temperature as their detectors,
they measured no signal at all. \ However, today experimenters have a great
advantage over the earlier researchers. \ By using the Casimir
effect,\cite{Casimir} it is possible to measure not only the excess
electromagnetic radiation arriving from a hot source, but indeed to measure
the entire spectrum of radiation surrounding an object. \ 

The Casimir effect involves the force between two uncharged conducting
parallel plates.\cite{Lam} \ The conducting boundary conditions at the plates
lead to forces associated with the radiation normal modes interacting with the
plates. \ From the magnitude of the force and from the dependence of the force
on the separation between the plates, it is possible to determine the entire
spectrum of random classical radiation surrounding the plates. \ At high
temperature $T$ or large separations $l$, the \textit{Casimir force expected
from the Rayleigh-Jeans spectrum of random classical radiation} surrounding
conducting plates of area $\mathcal{A}$ and separation $l$ is\cite{B1974c}
\begin{equation}
F_{RJ}=-\frac{\zeta(3)k_{B}T\mathcal{A}}{4\pi l^{3}}%
\end{equation}
At low temperatures, this force falls to zero along with the temperature $T.$
\ However, experimental measurements\cite{Cmes} of the Casimir force show that
the force between the plates does not vanish with vanishing temperature
$T\rightarrow0,$ but rather goes to the form which is temperature independent,%
\begin{equation}
F_{zp}=-\frac{\pi^{2}\hbar c\mathcal{A}}{240l^{4}} \label{Casimir2}%
\end{equation}
where $\hbar$ is a constant which must be fitted from experiment, and which
indeed takes the same numerical value as Planck's constant.\cite{hbar}
\ \textit{Interpreting the zero-temperature Casimir force within classical
electromagnetic theory},\cite{B1974c} we conclude that, surrounding the
conducting plates, there must be a spectrum of random classical radiation
corresponding to an average energy per normal mode%
\begin{equation}
U_{zp}(\omega)=\frac{1}{2}\hbar\omega. \label{zpe2}%
\end{equation}
This is the classical electromagnetic zero-point radiation of which the
physicists of a century ago were unaware. \ The experimentalist of that
earlier period were unable to measure this temperature-independent random
radiation, and, during the crucial period of decision, the theoretical
physicists of that earlier era did not anticipate the possibility of such
zero-point radiation. \ 

Of course today, the Casimir force is usually interpreted through quantum
theory, and some physicist wish to claim that classical zero-point energy
cannot be used within a classical electromagnetic theory.\cite{roles}
\ However, the classical electromagnetic calculations for the Casimir forces
are perfectly valid classical calculations.\cite{B1974c} \ Indeed, the
physicists at the turn of the 20th century treated thermal radiation as random
\textit{classical} radiation.

\subsection{Properties of Classical Zero-Point Radiation}

The zero-point radiation spectrum measured in Casimir experiments and
appearing in Eq. (\ref{zpe2}) leads to an energy spectrum%
\begin{equation}
\rho_{zp}(\omega)=[\omega^{2}/(\pi^{2}c^{3})]U_{zp}(\omega),
\end{equation}
where the factor $[\omega^{2}/(\pi^{2}c^{3})]$ is the number of normal modes
per unit (angular) frequency interval, and leads to a divergent energy density%
\begin{equation}
u_{zp}=%
{\textstyle\int_{0}^{\infty}}
d\omega\rho_{zp}(\omega)\rightarrow\infty.
\end{equation}
\ Despite the divergence of the energy density $u_{zp}$, we anticipate no
electromagnetic problems because each electromagnetic system interacts with
radiation within only a limited range of frequencies. \ 

At thermal equilibrium at positive temperature $T>0$, we expect the thermal
energy spectrum $\rho_{T}(\omega,T)$ to be in addition to the zero-point
energy $\rho_{zp}(\omega)$ which exists at zero temperature. $\ $The total
spectrum is the sum $\rho(\omega,T)=\rho_{T}(\omega,T)+\rho_{zp}(\omega),$ or,
in terms of the average energy per normal mode,%
\begin{equation}
U(\omega,T)=U_{T}(\omega,T)+U_{zp}(\omega),
\end{equation}
where $U(\omega,T)$ is the total electromagnetic energy in the radiation mode.
\ The energy density $u_{T}$ due to thermal radiation $\rho_{T}(\omega,T)$ is
indeed finite%
\begin{equation}
u_{T}=%
{\textstyle\int_{0}^{\infty}}
d\omega\rho_{T}(\omega)<\infty. \label{utherm}%
\end{equation}
When we remove all possible thermal radiation from a container by going to
zero temperature, what is left is the zero-point radiation.

Crucially, the spectrum of \textit{zero-point} radiation appearing in Eq.
(\ref{zpe2}) is \textit{Lorentz invariant}; it takes the same spectral form in
any inertial frame.\cite{Mar}\cite{B1969b} \ Indeed, only a spectrum leading
to a divergent energy density can look the same in every inertial frame. \ On
the other hand, \textit{thermal} radiation above the zero-point radiation has
a finite energy density and a preferred inertial frame; the preferred frame is
that of the container in which the radiation is at equilibrium.

It will be useful in the subsequent analysis to deal with the $\sigma
_{ltU^{-1}}$-scale invariance of classical electromagnetism which leaves
invariant the fundamental constants $c,~e,$ and $\hbar.$ \ In addition to
being Lorentz-invariant, the spectrum of classical zero-point radiation in Eq.
(\ref{zpe2}) is also $\sigma_{ltU^{-1}}$-scale invariant.\cite{scale} \ \ By
this we mean that if all lengths are transformed by the multiplicative factor
$\sigma$ so that $l\rightarrow l^{\prime}=\sigma l,$ while all times are
transformed as $t\rightarrow t^{\prime}=\sigma t,$ and all energies are
transformed as $U\rightarrow U^{\prime}=\sigma^{-1}U,$ then the random
zero-point radiation is unchanged. \ Indeed, the spectrum is unchanged, since,
$U_{zp}^{\prime}(\omega^{\prime})=U_{zp}(\omega)/\sigma=(1/2)\hbar
\omega/\sigma=(1/2)\hbar\omega^{\prime}.$ \ Under an adiabatic expansion or
compression of the thermal radiation in a spherical conducting-walled cavity,
the total spectrum is transformed, but remains a blackbody spectrum at a new
temperature. \ However, there is no mixing of the average zero-point and
thermal contributions during the adiabatic change of the modes' frequency,
wavelength, and energy; the zero-point spectrum is $\sigma_{ltU^{-1}}%
$-invariant and $U_{zp}(\omega)$ is mapped onto itself, $U_{zp}(\omega
)\rightarrow U^{\prime}(\omega^{\prime})=U_{zp}(\omega^{\prime}),$ while the
thermal radiation $U_{T}(\omega,T)$ is mapped onto thermal radiation at a new
scale-transformed temperature $U_{T}(\omega,T)\rightarrow U_{T}(\omega
,T^{\prime}),~$where $T^{\prime}=T/\sigma$. \ 

\subsection{How is Zero-Point Radiation Different from Thermal Radiation?}

One may ask, \textquotedblleft What is the difference between zero-point
radiation and thermal radiation within classical physics?\textquotedblright%
\ \ The answer is that there is no difference at all, except for the spectrum
of the radiation. \ Thus for classical radiation in equilibrium in an
enclosure,\ if we are told the average energy of a radiation mode without
being told its frequency, then we do not know how much of the average energy
is zero-point energy and how much is thermal energy. \ However, the spectrum
of zero-point radiation is $\sigma_{ltU^{-1}}$-scale invariant so that the
ratio $U_{zp}(\omega)/\omega$ between the average energy of the radiation mode
and the frequency is the same for all modes, $U_{zp}(\omega)/\omega
=U_{zp}(\omega^{\prime})/\omega^{\prime}.$ \ On the other hand, for positive
temperature $T>0,$ the ratios involving the total energy $U$ of the radiation
modes are related as $U(\omega,T)/\omega>U(\omega^{\prime},T)/\omega^{\prime}$
for $\omega\,<\omega^{\prime}$ and $T>0,$ since the thermal contribution to
the total energy of a radiation mode decreases with increasing frequency. \ In
thermal equilibrium, the radiation energy above the zero-point energy is
thermal energy. \ By going to sufficiently high frequency where the thermal
energy contribution becomes ever smaller, we can determine the zero-point
radiation spectrum which underlies the thermal contribution. \ 

This classical idea that both zero-point radiation and thermal radiation are
part of a single spectrum of random radiation is quite different from the
prevailing quantum view that quantum zero-point energy (involving no photons)
is quite different from the photons existing at positive temperature. \ No
such distinction exists within classical physics, nor in the measurements of
Casimir forces.

\section{Relativistic Mechanical Systems for Use in Relativistic
Electrodynamics}

We have stressed that the misapprehension that classical physics leads
inevitably to the Rayleigh-Jeans spectrum for thermal equilibrium arises
because of the erroneous use of nonrelativistic mechanical systems as agents
of radiation equilibrium for not only the low-frequency portion of the
spectrum, but for the entire spectrum. \ In order to understand classical
radiation equilibrium, we must demand that the mechanical agents of
equilibrium are compatible with relativistic electrodynamics. \ In the present
article, we will consider only relativistic point charges in Coulomb
potentials (regarded as part of relativistic classical electrodynamics), and
point charges in harmonic-oscillator potentials \textit{in the zero-amplitude
limit}. \ The harmonic oscillator system appears frequently in historical
accounts and can be regarded as the low-velocity limit of a relativistic
mechanical system. \ 

Also, it is convenient to introduce the action variables $J_{i}$ which appear
in graduate courses in mechanics. \ In the early 20th century, the action
variables appeared in Bohr-Sommerfeld quantization and played a role in
Ehrenfest's adiabatic theorem. \ These variables are also ideal parameters for
use in thermodynamic systems. \ The action variables $J_{i}$ are both
adiabatic invariants and also $\sigma_{ltU^{-1}}$-scale invariants. \ The
dimensions of the action variables are $energy\times time,$ and hence are
invariant under a $\sigma_{ltU^{-1}}$-scaling transformation.\cite{quant} \ 

\subsection{Harmonic Oscillator in the Point Limit}

A point charge in a harmonic-oscillator potential can be fitted into
relativistic electrodynamics in the limit of zero size for the oscillation
excursion. \ The harmonic oscillator is treated in the same fashion as the
electromagnetic radiation modes, except that the spatial extent of the
mechanical oscillator is taken as vanishingly small. \ In this case, the
energy $\mathcal{E}_{osc}$ of a one-dimensional harmonic oscillator
$\mathcal{E}_{osc}=m\dot{x}^{2}/2+\kappa x^{2}/2=m\dot{x}^{2}/2+m\omega
_{0}^{2}x^{2}/2$ can be given in terms of action-angle variables
as\cite{Goldstein}
\begin{equation}
\mathcal{E}_{osc}(\omega_{0},J)=J\omega_{0}. \label{Eosc}%
\end{equation}
The mass $m$ and spring-constant $\kappa$ are not discussed separately from
the characteristic frequency $\omega_{0}=(\kappa/m)^{1/2},$ while the
displacement $x$ and velocity $v$ are regarded as so small as to be
negligible. \ The system is $\sigma_{ltU^{-1}}$-scale covariant because energy
and time are connected as in Eq. (\ref{Eosc}) while lengths do not appear.
\ As pointed out by Planck, the point dipole oscillator in random classical
radiation acquires an average energy $U_{osc}(\omega_{0})=\left\langle
\mathcal{E}_{osc}\mathcal{(\omega}_{0})\right\rangle $ equal to the average
energy $U(\omega)=\left\langle \mathcal{E(\omega)}\right\rangle $ of the
radiation normal modes at the same frequency $\omega=\omega_{0}$ as the
oscillator.\cite{Planck}\cite{Lavenda} \ Also, the probability distribution
for the action variable $J$ of the mechanical harmonic oscillator is the same
as that for the radiation modes at the same frequency as the oscillator
frequency. \ 

\subsection{Point Charge in a Coulomb Potential}

The far more important relativistic system is that of a point charge in a
Coulomb potential, since, in relativistic classical electrodynamics, point
charges interact through electromagnetic fields . \ The relativistic energy
$\mathcal{E}_{C}=m\gamma c^{2}-Ze^{2}/r$ (with $\gamma=(1-v^{2}/c^{2}%
)^{-1/2})$ of a point charge $e$ in a Coulomb potential $V_{C}(r)=-Ze^{2}/r$
can be written in terms of action variables as \cite{Goldstein}%
\begin{equation}
\mathcal{E}_{C}(m,J_{2},J_{3})=mc^{2}\left[  1+\left(  \frac{Ze^{2}/c}%
{J_{3}-J_{2}+\sqrt{J_{2}^{2}-(Ze^{2}/c)^{2}}}\right)  ^{2}\right]  ^{-1/2},
\label{CoulE}%
\end{equation}
where $J_{1}=J_{\phi},~J_{2}=J_{\phi}+J_{\theta},~$and $J_{3}=J_{\phi
}+J_{\theta}+J_{r}$. \ We notice that the energy $\mathcal{E}_{C}%
(m,J_{2},J_{3})$ is a product of the particle rest energy $mc^{2}$ and a
dimensionless function of $J_{i}/(Ze^{2}/c).$ \ This system is $\sigma
_{ltU^{-1}}$-scale covariant with the mass $m$ as the one scaling parameter.
\ The quantities $J_{i},~Z,$~$e,$ and $c$ are all $\sigma_{ltU^{-1}}%
$-invariant. \ All lengths, times, and energies can be found from the
fundamental length $e^{2}/(mc^{2}),$ fundamental time $e^{2}/(mc^{3}),$ and
fundamental energy $mc^{2},$ multiplied by a function of $J_{i}/(Ze^{2}/c).$
\ The existence of two different regimes associated with relativistic particle
behavior is vividly illustrated in the unfamiliar trajectories of relativistic
particles in a Coulomb potential, which can be strikingly different from the
nonrelativistic orbits associated with conic sections.\cite{B2004}

For the case of a circular orbit\cite{B2004} where $J_{r}=0$ and $J_{2}%
=J_{3}=J,$ the energy becomes $\mathcal{E}_{C}(m,J)=mc^{2}\{1-[Ze^{2}%
/(Jc)]^{2}\}^{1/2}$ with velocity $v=Ze^{2}/J$ and the velocity ratio becomes
$v/c=Ze^{2}/(Jc).$ \ We notice that the range of $J_{2}=J_{3}=J$ is limited,
$Ze^{2}/c<J<\infty,$ corresponding to the particle velocity which is less than
$c.$ \ 

\section{Thermal Behavior in Terms of Action-Angle Variables in Relativistic
Classical Electrodynamics}

Having mentioned some aspects of \textit{relativistic} classical systems, we
now turn to the \textit{thermal} behavior of classical systems.

\subsection{Probability Distribution for Disordered Systems}

The classical thermodynamics of both particles and radiation can be treated as
involving random magnitudes for the action variables and random phases for the
angle variables. \ When a periodic system is discussed in terms of
action-angle variables, the system has an energy expression containing
mechanical parameters characterizing the intrinsic system itself (such as mass
$m$ or radiation-mode frequency $\omega$) and also the action variables
$J_{i}$\ which characterize the particular state of the system (such as the
system's angular momentum or energy). \ An ensemble of identical mechanical
systems with the same mechanical parameters and with differing energies will
be described by a probability distribution in the action variables $J_{i},$
analogous to the distribution on phase space used in statistical
thermodynamics. $\ \ $For example, a one-spatial-degree-of-freedom system in
thermal equilibrium with a heat bath at temperature $T$ has a probability
distribution $P(J,T)dJ$ associated with the randomness of one action variable
$J.$ \ The average energy $U(T)$ of the system is found by integrating the
energy $\mathcal{E(}J)$ over the probability distribution $P(J,T)dJ$ for the
action variable%
\begin{equation}
U(T)=%
{\textstyle\int}
dJ\,\mathcal{E(}J)\,P(J,T).
\end{equation}

\subsection{Probability Distribution for Radiation Modes}

For a radiation normal mode of frequency $\omega$ in an enclosure at
temperature $T$, the quantity $P(\omega,J,T)dJ$ can be determined by the
fundamental properties of waves. \ Within classical physics, the fluctuations
for waves can be described in terms of interference between waves of differing
frequency, leading to a probability distribution for random wave behavior at
frequency $\omega$ in the form of a Gaussian distribution in the wave
amplitude, which corresponds to an exponential distribution in the action
variable.\cite{Rice} \ Thus the probability distribution for a radiation mode
in a thermal bath involves the mode energy $\mathcal{E(}J)=J\omega$ divided by
the average mode energy $U(\omega,T)$ and takes the form\cite{B1978c}
\begin{equation}
P(\omega,J,T)dJ=\exp\left[  \frac{-J\omega}{U(\omega,T)}\right]  \frac{\omega
}{U(\omega,T)}dJ. \label{probJT}%
\end{equation}
The distribution is normalized so that $%
{\textstyle\int_{0}^{\infty}}
dJ\,P(\omega,J,T)=1.$ \ It is from this probability distribution that the
entropy of the system would be evaluated, if the entropy function were known.
\ We notice that the probability distribution satisfies $\sigma_{ltU^{-1}}%
$-scale covariance for the theory, but the distribution (\ref{probJT}) is not
$\sigma_{ltU^{-1}}$-scale invariant, since, under a $\sigma_{ltU^{-1}}$-scale
transformation, the energy is mapped to a new energy at a new temperature. \ 

In the case of classical zero-point radiation, we know the explicit form for
the average energy from Casimir force measurements, namely $U(\omega
,0)=(1/2)\hbar\omega.$ Therefore equation (\ref{probJT}) becomes%
\begin{equation}
P(\omega,J,0)dJ=\exp\left[  \frac{-J\omega}{(1/2)\hbar\omega}\right]
\frac{\omega}{(1/2)\hbar\omega}dJ=\exp\left[  \frac{-J}{\hbar/2}\right]
\frac{2}{\hbar}dJ. \label{probJ00}%
\end{equation}
Thus at zero temperature, the zero-point radiation probability distribution
for the action variable $J$ depends upon Planck's constant $\hbar,~$and is
exactly the same for every radiation mode, independent of the frequency
$\omega$ of the mode. \ The distribution (\ref{probJ00}) is $\sigma_{ltU^{-1}%
}$-scale \textit{invariant} since $J$ and $\hbar$ are each $\sigma_{ltU^{-1}}%
$-scale invariant.

During an adiabatic change for a system, the distribution of the action
variables $J_{i}$ remains unchanged, since the $J_{i}$ are adiabatic
invariants. \ On the other hand, when heat energy is added to a system, the
transfer of heat energy without work involves a change in the distribution of
action variables while the physical dimensional parameters of the system
remain unchanged. \ At zero temperature, there can be no transfer of heat and
therefore no change in the distribution of action variables, even when the
physical dimensional parameters of the system are changed and work is done.
\ We notice in Eq. (\ref{probJ00}) (which holds at $T=0)$ that indeed the
distribution of the action variable $J$ for a radiation mode of frequency
$\omega$ does not change when the frequency of the mode is changed. \ 

\subsection{Probability Distribution for a Point Charge in a Coulomb
Potential}

We have mentioned two possible simple mechanical systems which can be regarded
as part of relativistic classical electrodynamics: the point dipole oscillator
(no spatial extent) and the point charge in a Coulomb potential. \ A
mechanical harmonic oscillator (taken in the point-size limit with negligible
velocity) is regarded as a passive mechanical system since it cannot change
the frequency spectrum of random radiation. \ It behaves exactly as a
radiation mode as concerns its distribution of action variables in thermal
radiation, and so involves no new information.\ 

The more interesting situation involves a relativistic point charge $e$ of
mass $m$ in a Coulomb potential $V_{C}(r)=-Ze^{2}/r$ at temperature $T$. \ In
this case, the probability distribution for the mechanical action variables
$J_{i}$ may depend upon such quantities as the mass $m$, the potential
constant $Ze^{2}$, the temperature $T$, and the speed of light $c$.
$\ $Because probability is a dimensionless quantity, the probability
distribution must depend upon dimensionless ratios, such as $\mathcal{E}%
_{C}(J_{i})/(k_{B}T)$ involving the system energy $\mathcal{E}_{C}(J_{i})$
given in Eq. (\ref{CoulE}) divided by $k_{B}T.$ \ However, we note from Eq.
(\ref{CoulE}) that the ratio $\mathcal{E}_{C}\mathcal{(}J_{i})/(k_{B}T)$ can
be simplified so as to involve a product of the ratio $mc^{2}/(k_{B}T)$ and a
function of $J_{i}/(Ze^{2}/c).$ \ Indeed, from dimensional considerations
alone, the functional form of the probability distribution for the $J_{i}$
must depend on the dimensionless ratios $J_{i}/(Ze^{2}/c)$ and $mc^{2}/T,$
giving a functional form $P_{C}[J_{i}/(Ze^{2}/c),mc^{2}/(k_{B}T)][c/(Ze^{2}%
)]dJ_{i}.$

We have noted that thermal radiation involves two different regimes depending
upon the ratio $\hbar\omega/\left(  k_{B}T\right)  $ determining whether
Lorentz-invariant zero-point radiation or thermal radiation provides the
dominant energy at a given frequency. The two-regime distinction between
zero-point energy and thermal energy for the relativistic particle in a
Coulomb potential is also clear. \ The distinction involves the ratio
$mc^{2}/(k_{B}T)$. \ This ratio reflects the ratio of mechanical frequency to
temperature, since (from the energy expression in Eq. (\ref{CoulE})) the
frequency of the orbital motion $\omega_{i}(J_{i})=\partial\mathcal{E}%
_{C}(J_{i})/\partial J_{i}$ is proportional to the mass $m$ of the charged
particle. \ The dominance of zero-point mechanical energy involves the
situation of large mass $m,$ where the frequency $\omega_{i}$ of orbital
motion is high, the ratio $mc^{2}/k_{B}T$ is large, and the contribution of
thermal energy is small. \ The high-frequency high-velocity limit of
mechanical energy is associated with relativistic zero-point energy, just as
the high-frequency limit of blackbody radiation is associated with
relativistic zero-point radiation. \ On the other hand, the thermal energy
becomes important in the opposite nonrelativistic limit where the mass $m$ is
small, the particle velocity is small, the orbital frequency $\omega$ is
small, and the ratio $mc^{2}/k_{B}T$ is small, in complete analogy with the
electromagnetic radiation situation where the thermal energy dominates at low
frequencies where the Rayleigh-Jeans form is indeed appropriate.

Crucially, at zero temperature, there is the possibility of non-zero energy
where the thermodynamic probability distribution $P_{C}[J_{i}/(Ze^{2}%
/c)][c/(Ze^{2})]dJ_{i}$ depends upon fundamental constants $Ze^{2}/c$ with no
dependence upon the mass $m$ because there is no dimensionless ratio available
which involves $m$. \ This $\sigma_{ltU^{-1}}$-scale-invariant situation is
exactly analogous to the situation for relativistic zero-point radiation
where, at zero temperature, the radiation modes have non-zero energy while the
thermodynamic probability distribution for the action variables (given in Eq.
(\ref{probJ00})) depends upon the fundamental constant $\hbar$ with no
dependence upon the frequency $\omega.$ \ The Coulomb potential of
relativistic classical electrodynamics can indeed support relativistic
zero-point energy through the fundamental constant $e^{2}/c$, just as
electromagnetic radiation can support relativistic zero-point radiation with
an amplitude given by the fundamental scale factor $\hbar.$

\subsection{Nonrelativistic Classical Theory Cannot Support Zero-Point Energy}

We emphasize that within classical theory, zero-point particle energy which is
compatible with relativistic zero-point radiation exists only within
\textit{relativistic} electrodynamics, where, for the Coulomb potential, large
mass is associated with high orbital frequency, and, for a circular
orbit,\cite{B2004} the velocity increases to $c$, $v\rightarrow c,$ as $J$
decreases to $Ze^{2}/c,$ $J\rightarrow Ze^{2}/c.$ \ On the other hand, in the
nonrelativistic mechanics of bounded potentials, large mass is ordinarily
associated with low frequency, and the fundamental velocity $c$ does not
appear. \ Now, for a point harmonic oscillator, the particle amplitude and
velocity can be regarded as negligible, and there is no conflict with a finite
value of $c.$ \ However, there is an obvious conflict for nonlinear
nonrelativistic mechanical systems where the particle amplitudes and
velocities cannot be regarded as zero. \ The assumptions ordinarily made for a
multiply periodic system with action variables $J_{i}$ in (nonrelativistic)
classical statistical mechanics are that\cite{vanVl} \textquotedblleft the
energy becomes infinite with each of the $J^{\prime}s,$ and . . . the
amplitudes and energy vanish with the $J^{\prime}s.$\textquotedblright\ \ This
suggests that the velocity of the particle of the system can increase without
limit as either the temperature of the system becomes large or the frequency
of the system becomes large when adiabatic work is done on the system. \ The
possibility of arbitrarily large system velocity is unacceptable in classical
electrodynamics; there is no way in which a charge moving with velocity in
excess of $c$ can be reconciled with classical radiation equilibrium.

\section{Thermodynamics of Blackbody Radiation}

Having reviewed some basic aspects for the current understanding of blackbody
radiation within relativistic classical physics, we now turn to a historical
perspective on blackbody radiation.

\subsection{Thermodynamics in the History of Blackbody Radiation}

Both thermodynamics and classical electrodynamics were developed during the
19th century.\cite{Kuhn} \ The theoretical treatment of blackbody radiation is
founded upon Kirchoff's blackbody analysis of 1860. \ The original
thermodynamics discussion involved the emissive and absorptive properties of
materials. \ Later the idea of cavity radiation was introduced. \ It was noted
that the blackbody energy density $u_{T}(T)$ and the blackbody radiation
spectrum $\rho_{T}(\omega,T)$ were universal functions independent of the
material forming the walls of the cavity enclosing the radiation. \ A single
\textquotedblleft black particle\textquotedblright\ which scattered radiation
within a conducting-walled cavity would change an arbitrary radiation
distribution over to the spectrum of thermal equilibrium. \ Thus a
\textquotedblleft black particle\textquotedblright\ encodes within itself and
its interactions the laws of thermodynamic equilibrium for radiation.

In 1879, based upon experimental measurements, Stefan suggested that the total
thermal radiation energy $\mathcal{U}_{T}$ in a container with volume
$\mathcal{V}$ at temperature $T$ was given by
\begin{equation}
\mathcal{U}_{T}=a_{S}\mathcal{V}T^{4} \label{SB}%
\end{equation}
where $a_{S}$ was constant. \ In 1884, Boltzmann derived this Stefan-Boltzmann
relation (\ref{SB}) by combining ideas from thermodynamics and classical
electrodynamics. \ With the appearance of the Stefan-Boltzmann law, blackbody
radiation was easily recognized to involve a new fundamental constant $a_{S}$
(Stefan's constant). \ Today Stefan's constant is restated in terms of more
familiar constants as\cite{Morse}%
\begin{equation}
a_{S}=\frac{\pi^{2}k_{B}^{4}}{15\hbar^{3}c^{3}}. \label{Stef}%
\end{equation}
Finally, in 1893, Wien applied thermodynamics to an adiabatic compression of
thermal radiation and derived the displacement law. \ This law indicated that
the blackbody radiation spectrum was of the form $\rho_{T}(\omega
,T)=const\times\omega^{3}f(\omega/T).$ \ Subsequent work showed that Wien's
displacement law was equivalent to the statement that the average energy per
normal mode of radiation was\ given by $U_{T}(\omega,T)=\omega f(\omega/T)$
where $f(\omega/T)$ was an unknown function. \ 

\subsection{Thermodynamics of a Single Radiation Mode}

Although Kirchoff, Stefan, Boltzmann, and Wien all thought in terms of thermal
radiation having a finite energy density, we now know that measurements of
Casimir forces indicate the existence of classical zero-point energy with its
divergent energy density. \ Cole \ has reviewed the thermodynamics of
blackbody radiation when zero-point radiation is included.\cite{Cole1992}
\ \ It also seems helpful to reconsider the thermodynamics associated with a
single radiation mode in order to clarify the situation in the presence of
zero-point radiation. \ 

The thermodynamics of an electromagnetic radiation mode (or indeed of a
harmonic oscillator) is particularly simple since it involves only two
thermodynamic variables: frequency $\omega$ and temperature $T$.\cite{B2003c}
\ In equilibrium with a heat bath, the average mode energy is denoted by
$U=\left\langle \mathcal{E}\right\rangle =\left\langle J\right\rangle \omega$
(where $J$ is the action variable), and satisfies the first law of
thermodynamics, $dQ=dU+dW,$ with the entropy $S$ satisfying $dS=dQ/T.$ \ Now
since $J$ is an adiabatic invariant, the work done by the system is given by
$dW=-\left\langle J\right\rangle d\omega=-(U/\omega)d\omega.$ \ Combing these
equations, we have $dS=dQ/T=[dU-(U/\omega)d\omega]/T.$ $\ $Writing the
differentials for $S$ and $U$ in terms of $T$ and $\omega,$ we have
$dS=(\partial S/\partial T)dT+(\partial S/\partial\omega)d\omega$ and
$dU=(\partial U/\partial T)dT+(\partial U/\partial\omega)d\omega.$ \ Therefore
$\partial S/\partial T=(\partial U/\partial T)/T$ and $\partial S/\partial
\omega=[(\partial U/\partial\omega)-(U/\omega)]/T.$ \ Now equating the mixed
second partial derivatives $\partial^{2}S/\partial T\partial\omega
=\partial^{2}S/\partial\omega\partial T,$ we have $(\partial^{2}%
U/\partial\omega\partial T)/T=-(\partial U/\partial T)/(\omega T)+(\partial
^{2}U/\partial T\partial\omega)/T+[(U/\omega)-(\partial U/\partial
\omega)]/T^{2}$ or $0=(\partial U/\partial T)/(\omega T)-[(U/\omega)-(\partial
U/\partial\omega)]/T^{2}.$ \ The general solutions of the differential
equations for $U$ and $S$ involve one unknown function of the single variable
$\omega/T.$ \ Indeed, the whole thermodynamic system can be described by the
unknown thermodynamic potential function\cite{Garrod} $\phi(\omega/T),$ a
function of one variable $\omega/T,$ where the energy is given by
\begin{equation}
U(\omega,T)=T^{2}(\partial\phi/\partial T)_{\omega}=-\omega\phi^{\prime
}(\omega/T) \label{Wien2}%
\end{equation}
and the entropy corresponds to
\begin{equation}
S(\omega/T)=\phi(\omega/T)+U(\omega,T)/T=\phi(\omega/T)-(\omega/T)\phi
^{\prime}(\omega/T). \label{entr2}%
\end{equation}
\ The result (\ref{Wien2}), obtained here purely from thermodynamics,
corresponds to the familiar Wien displacement law of classical physics.

\subsection{Thermodynamic Limits: Zero-Point Radiation and Equipartition}

Although the thermodynamic analysis culminating in the Wien displacement
theorem of 1893 simplified the classical blackbody problem down to an unknown
function of one variable $\omega/T$, the blackbody spectrum was still
undetermined. \ However, it is noteworthy that the unknown thermodynamic
potential function $\phi(\omega/T)$ allows two natural limits which make the
energy $U(\omega,T)$ independent of one of its variables.\cite{B2003c} \ 

\subsubsection{Equipartition Limit at High Temperature}

If the function $\phi(\omega/T)\rightarrow-const_{1}\times\ln(\omega/T)$ for
small arguments $\omega/T<<1$, then the average energy $U(\omega,T)$ becomes
independent of $\omega,$ $U(\omega,T)=-\omega\phi^{\prime}(\omega
/T)\rightarrow\omega\times const_{1}\times(T/\omega)=T\times const_{1}.$
\ Choosing the constant $const_{1}$ as Boltzmann's constant $k_{B},~$this
limit becomes
\begin{equation}
U(\omega,T)\rightarrow k_{B}T\text{ \ \ for \ large }T. \label{LT}%
\end{equation}
This is the familiar equipartition result of nonrelativistic kinetic theory
and nonrelativistic classical statistical mechanics. \ It agrees with the
experimentally-measured thermal radiation at high temperatures and low frequencies.

\subsubsection{Zero-Point Energy Limit at Low Temperature}

On the other hand, if the function $\phi(\omega/T)\rightarrow-const_{2}%
\times(\omega/T)$ for large arguments $\omega/T>>1,$ then the average energy
$U(\omega,T)$ becomes independent of the temperature $T,$ $U(\omega
,T)=\omega\times const_{2}.$ \ This energy corresponds to relativistic
zero-point energy for the radiation mode provided that the constant
$const_{2}$ is taken as half Planck's constant $\hbar/2,$%
\begin{equation}
U(\omega,T)\rightarrow(1/2)\hbar\omega\text{ \ \ for small }T\text{.}
\label{ST}%
\end{equation}
This corresponds to the relativistic-invariant and scale-invariant zero-point
radiation which is measured experimentally from Casimir forces. \ 

\subsubsection{Zero-Point Radiation Has Zero Entropy}

We see here that zero-point radiation fits in naturally with the Wien
displacement theorem for thermal radiation. \ Furthermore, the entropy
associated with zero-point radiation vanishes, since, for $\phi(\omega
/T)=-(\hbar/2)(\omega/T),$ the entropy following from Eq. (\ref{entr2}) is
\begin{equation}
S(\omega,T)=\phi(\omega/T)-(\omega/T)\phi^{\prime}(\omega/T)=-(\hbar
/2)(\omega/T)-(\omega/T)(-\hbar/2)=0.
\end{equation}
Thus zero-point radiation makes no contribution to the thermodynamic entropy
$S(\omega,T)$. \ The derivation of the Stefan-Boltzmann law in Eq. (\ref{SB})
involves only the total thermal energy $\mathcal{U}_{T}(T)$ energy obtained by
summing over the mode thermal energies $U_{T}(\omega,T)$ associated with
changes in thermal mode entropies $S(\omega,T)$, and so refers to the thermal
radiation energy above the zero-point radiation energy, $U_{T}(\omega
,T)=U(\omega,T)-U(\omega,0)$. \ 

\subsection{The Planck Spectrum is the \textquotedblleft Smoothest
Interpolation\textquotedblright\ Between Energy Equipartition and Zero-Point
Energy}

The thermodynamic potential can be rewritten in terms of the constants $k_{B}$
and $\hbar$ appearing in Eqs. (\ref{LT})\ and (\ref{ST}) as a function
$\phi\lbrack\hbar\omega/(k_{B}T)]$ of the single dimensionless variable
$\hbar\omega/(k_{B}T)$ with asymptotic forms%
\begin{align}
\phi\lbrack\hbar\omega/(k_{B}T)]  &  \rightarrow-k_{B}\ln(\hbar\omega
/k_{B}T)\text{ \ for \ }\hbar\omega/k_{B}T<<1,\nonumber\\
\phi\lbrack\hbar\omega/(k_{B}T)]  &  \rightarrow-k_{B}\frac{1}{2}\frac
{\hbar\omega}{k_{B}T}\text{ \ for \ }\hbar\omega/k_{B}T>>1,
\end{align}
where the energy relation (\ref{Wien2}) now becomes%
\begin{equation}
U(\omega,T)=-\frac{\hbar\omega}{k_{B}}\phi^{\prime}[\hbar\omega/(k_{B}T)].
\label{Wien4}%
\end{equation}

The physicists of the 1890s were unaware of zero-point radiation and so did
not discuss the asymptotic limits of the thermodynamic potential. \ Thus the
physicists of this earlier era moved on to aspects which did not involve
thermodynamics. \ However, since we are now working at a time when the
asymptotic limits are known, and also we are used to the basic idea of
thermodynamics involving smooth functions, we can try to evaluate the full
blackbody radiation spectrum by making the \textquotedblleft smoothest
possible\textquotedblright\ interpolation\ between the known asymptotic
limits.\cite{B2003c}

Now the idea of a \textquotedblleft smoothest possible\textquotedblright%
\ interpolation seems ambiguous except in the case where we are told that a
function $f(z)$ takes the value $a$ at$~z=0,$ has first derivative $-b$ at
$z=0,$ $(a,b>0),~$and goes to zero as $z$ goes to infinity. \ In this case, it
seems clear that the smoothest possible interpolation from $f(z)\approx a-bz$
near $z=0$ to $f(\infty)=0$ involves $f(z)=a\exp[-zb/a],$ since this function
meets all the asymptotic requirements, introduces no new parameters or special
points, and has all derivatives related back to the function itself. \ 

Evaluation of the \textquotedblleft smoothest possible\textquotedblright%
\ interpolation can be carried out as follows. \ First we rescale the
thermodynamic potential by taking $\phi\lbrack\hbar\omega/(k_{B}T)]/k_{B} $
and work with the scale-invariant $\phi(z)$ where $z=$ $\hbar\omega/(k_{B}T).$
\ Next we remove the logarithmic behavior by taking the exponential of
$-\phi(z).$ \ Now we try to arrange for the difference between $\exp
[-\phi(z)]$ and some known function to have exactly the asymptotic properties
of a \textquotedblleft smoothest\textquotedblright\ function of the form
$a\exp[-zb/a].$ \ Clearly the required form appears for the combination
$\exp[z/2]-\exp[-\phi(z)],$ since we have the asymptotic limits%
\begin{align}
\exp[z/2]-\exp[-\phi(z)]  &  \rightarrow1+z/2-z=1-z/2\text{ \ \ for
\ }z\rightarrow0,\nonumber\\
\exp[z/2]-\exp[-\phi(z)]  &  \rightarrow0\text{ \ \ for \ }z\rightarrow\infty.
\end{align}
Thus the \textquotedblleft smoothest possible\textquotedblright\ interpolation
between the thermodynamic limits is $\exp[z/2]-\exp[-\phi(z)]=\exp[-z/2],$ so
that $\exp[-\phi(z)]=\exp[z/2]-\exp[-z/2]=2\sinh[z/2]$. \ Now restoring the
scale for $\phi$ and, inserting $z=\hbar\omega/(k_{B}T),$ we have%
\begin{equation}
\phi\lbrack\hbar\omega/(k_{B}T)]=-k_{B}\ln\{2\sinh[\hbar\omega/(2k_{B}T)]\}.
\end{equation}
But then the average energy per normal mode in Eq. (\ref{Wien4}) is
\begin{equation}
U_{Pzp}(\omega,T)=-\frac{\hbar\omega}{k_{B}}\phi^{\prime}[\hbar\omega
/(k_{B}T)]=\frac{1}{2}\hbar\omega\coth\left(  \frac{\hbar\omega}{2k_{B}%
T}\right)  =\frac{\hbar\omega}{\exp[\hbar\omega/(k_{B}T)]-1}+\frac{1}{2}%
\hbar\omega\label{Pl3}%
\end{equation}
which is precisely the Planck spectrum including zero-point energy. \ Thus an
easy interpolation between the thermodynamic limits (including the zero-point
energy limit) gives the experimentally-measured spectrum.

\section{Determining the Unknown Spectral Function: Historical Aspects}

Historically, the thermodynamic analysis of blackbody radiation reached its
limit in the Wien displacement theorem. \ At the turn of the 20th century,
there was no thought of relativistic zero-point radiation, and hence there was
no attempt at any interpolation between the zero-point radiation and
equipartition asymptotic forms. \ What was recognized was that electromagnetic
radiation within a conducting-walled enclosure would not bring itself to
thermal equilibrium. \ Therefore it was the interaction of matter with
radiation in the sense of Kirchoff's \textquotedblleft black
particles\textquotedblright\ which brought about thermal radiation
equilibrium. \ Since the physics of matter was described in terms of classical
mechanics, radiation equilibrium somehow involved the interaction of radiation
with mechanical systems.

\subsection{Planck's Resonators}

At \ the end of the 19th century, it was Planck\cite{Planck} who introduced
linear electromagnetic \textquotedblleft resonators\textquotedblright\ in
hopes that they would serve as black particles and so determine the blackbody
radiation spectrum theoretically. \ Today we think of Planck's resonators in
terms of point electric dipole oscillators interacting with the surrounding
radiation.\cite{Lavenda}\cite{B2016c} \ Planck found that the average energy
$U_{osc}(\omega_{0},T)$\ of an oscillator of natural frequency $\omega_{0}$ in
random radiation at temperature $T$ is the same as the average energy
$U(\omega,T)$ per normal mode of the random radiation in the modes at the same
frequency $\omega=\omega_{0}$ as the natural frequency of the oscillator.
\ Here was a first equilibrium connection between matter and radiation.
\ However, despite Planck's original hope, a harmonic oscillator does not act
as a black particle. \ The average energy of the oscillator \textit{matches}
the average energy of the random radiation modes at $\omega_{0},\,\ $but the
oscillator does \textit{not determine} the spectrum of blackbody radiation.
\ The charged dipole oscillator scatters radiation so as to make the random
radiation spectrum more nearly isotropic, but the frequency spectrum of the
scattered radiation is exactly the same as the frequency spectrum of the
incident radiation.\cite{B1975a} \ \ 

\subsection{Conflict Between Classical Mechanics and Electrodynamics}

Even during the 19th century, it was recognized that classical mechanics and
classical electrodynamics had different, indeed clashing aspects. \ By the
turn of the 20th century, \textquotedblleft the relationship \ between
electrodynamics and mechanics had, for a generation, been growing increasingly
problematic.\textquotedblright\cite{Kuhn112} \ Classical mechanics has no
fundamental velocity, whereas classical electrodynamics incorporates a
fundamental speed $c$ corresponding to the speed of light in vacuum.
\ Furthermore, classical statistical mechanics (which is based upon
nonrelativistic classical mechanics) leads to the idea of kinetic energy
equipartition though point collisions of particles, whereas classical
electrodynamics does not consider point particle collisions but rather
involves long-range Coulomb forces which do not fit into nonrelativistic
classical statistical mechanics. \ Sharing energy by point collisions is
satisfactory for nonrelativistic mechanics but becomes dubious for a
relativistic theory with radiation.

In addition to the speed of light $c$, classical electrodynamics was found to
involve two new fundamental constants. \ Stefan's constant was introduced in
1879, and the electron charge $e$ was found around 1897. \ Thus the Coulomb
force involved a smallest characteristic charge $e,$ corresponding to the
electron charge. \ The conflicts between nonrelativistic classical mechanics
and classical electrodynamics became intense in the last years of the 19th
century when increasingly accurate experiments searched for the inertial frame
of the electrical ether, and also when increasingly accurate experiments
measured the heat transfer of blackbody radiation. \ Michelson and Morley
carried out what is today the most famous of the ether-search experiments, and
Paschen, Lummer and Pringsheim, Rubens and Kurlbaum provided the most famous
blackbody measurements. \ The conflict between the theories of mechanics and
electrodynamics reached a climax in the early years of the 20th century. \ It
was realized that classical electrodynamics was a relativistic theory
satisfying Lorentz transformations whereas Newtonian classical mechanics
satisfied Galilean transformations between inertial frames. \ \textit{However,
in the historical accounts of early 20th century physics, there appears to be
no hint that relativity might have any relevance for the problems of thermal
radiation.}\cite{Kuhn}

\subsection{Experimental Measurements in the Long-Wavelength Region}

In 1896, shortly after his derivation of the displacement law, Wien suggested
the expression for the blackbody spectrum $\rho_{W}(\omega,T)=\alpha\omega
^{3}\exp[-\beta\omega/T]$ \ $(\alpha,\beta$ positive constants) based on the
current experimental data and some vague ideas from the Maxwell velocity
distribution of classical particles in thermal equilibrium.\cite{Kuhn}
\ However, new experimental measurements in the long-wavelength (low
frequency) region by Lummer and Pringsheim and by Rubens and Kurlbaum in 1899
clearly disagreed with Wien's suggestion, although Wien's distribution
continued to represent the high-frequency region well. \ Informed of this
experimental disagreement, Planck attempted an interpolation between the new
experimentally-determined low-frequency spectrum and the satisfactory
high-frequency behavior appearing in Wien's distribution. \ The interpolation
involved the energy $U$ and entropy $S$ of the radiation modes. \ Planck's
interpolation of 1900 led to the average energy per normal mode%
\begin{equation}
U_{P}(\omega,T)=\frac{\hbar\omega}{\exp[\hbar\omega/(k_{B}T)]-1}. \label{Pl2}%
\end{equation}
Planck's radiation spectrum in Eq. (\ref{Pl2}) fitted the experimental
measurements of thermal radiation so well that it became a focus of
theoretical analysis thereafter.

\subsection{Direct Use of the Equipartition Theorem of Classical Statistical
Mechanics}

In the early years of the 20th century, Rayleigh and Jeans gave the derivation
of the classical radiation spectrum which is still quoted in all the textbooks
of modern physics. \ In 1900, Rayleigh suggested that the \textquotedblleft
Maxwell-Boltzmann doctrine of the partition of energy\textquotedblright\ be
applied to the thermal radiation in a cavity. \ He came up with the average
energy per radiation normal mode $U(\omega,T)=k_{B}T$. \ Since this spectrum
clearly did not fit the high-frequency part of the spectrum, he introduced an
exponential factor which cut off the spectrum at high frequency. \ In 1905,
Jeans extended the equipartition analysis on a carefully-argued basis, so that
today the radiation energy per normal mode%
\begin{equation}
U_{RJ}(\omega,T)=k_{B}T \label{equi}%
\end{equation}
is known as the Rayleigh-Jeans spectrum. \ This spectrum agreed with
experimental measurements in the low-frequency region, but was clearly wrong
at high frequency. \ Jeans felt so strongly about the firm theoretical basis
for his equipartition analysis that he suggested that the high-frequency
regions of the experimentally-measured spectrum might not correspond to
thermal equilibrium. \ 

It is noteworthy that Planck's constant does not appear in the Rayleigh-Jeans
spectrum (\ref{equi}). \ Rather, the Rayleigh-Jeans spectrum allows a
completely independent scaling in energy, unconnected with other quantities,
as is typical in nonrelativistic mechanics.

\subsection{Lorentz's Use of Maxwell's Velocity Distribution from Classical
Kinetic Theory}

In 1903, Lorentz gave a derivation\cite{Lor1903} of the Rayleigh-Jeans law
from classical physics based upon the interaction of radiation with the
electrons in a thin slab of material. \ The motion of the electrons was
described in terms of the Drude model involving free-particle motion with
collisions where the velocity distribution for the electrons was taken as the
Maxwell velocity distribution of nonrelativistic classical kinetic theory.
\ The approximations applied were all appropriate for long wave-length
radiation, and indeed Lorentz arrived at exactly the Rayleigh-Jeans result in
Eq. (\ref{equi}). \ Lorentz looked upon his work as a confirmation of the
understanding of the long-wavelength (low-frequency) portion of the spectrum. \ 

\subsection{Climax of the Classical Statistical Mechanical Arguments}

Although the use of classical statistical mechanical arguments to arrive at
the Rayleigh-Jeans spectrum is repeated in all the modern physics texts today,
the physicists of the first decade of the 20th century did not regard these
arguments as compelling. \ The classical statistical mechanical arguments were
regarded as merely providing an understanding for the long-wavelength region
of the blackbody spectrum. \ 

It was Lorentz's speech in Rome in 1908 which precipitated the change in the
accepted views regarding blackbody radiation. \ Lorentz reviewed the
derivations of the Rayleigh-Jeans spectrum from classical statistical
mechanics, and then raised the possibility that Jeans' view was correct;
\ perhaps the experimentally-measured high-frequency portions of the spectrum
did not represent thermal equilibrium. \ The experimentalists, who had
previously ignored Jeans' suggestions regarding the failure of thermal
equilibrium, did not ignore Lorentz's remarks; they reacted with outraged
ridicule to the idea that their careful measurements did not correspond to
thermal equilibrium. \ Lorentz then quickly retreated. \ \textquotedblleft
Only after Lorentz's Rome lecture does the physics profession at large seem to
have been confronted by what shortly came to be called the ultraviolet
catastrophe...\textquotedblright\cite{Kuhn195}

\section{The Place for Planck's Constant in Classical Physics}

In the reanalysis of the theoretical situation following Lorentz's speech and
the experimenters' contradictions, attention focused on the apparent absence
of any place for Planck's constant $\hbar$ within classical theory. \ 

\subsection{Blackbody Radiation Introduces a New Constant into Physics}

The experimental measurements of blackbody radiation indicated clearly that
the radiation involved two constants, not simply one. \ Indeed, Wien's
suggestion for the blackbody radiation spectrum $\rho_{W}(\omega
,T)=\alpha\omega^{3}\exp[-\beta\omega/T]$ involved two constants from fitting
the experimental data. \ Planck's work (relating the radiation spectrum $\rho$
to the average energy $U$ for an oscillator) gave for Wien's suggested
spectrum $\rho_{W}(\omega,T)$ the average oscillator energy $U_{W}%
(\omega,T)=\hbar\omega\exp[-\beta\omega/T]$ where Planck introduced the
constant $h=2\pi\hbar$. \ In May 1899, Planck determined the value of the
constant as $\ h=2\pi\hbar=6.885\times10^{-27}$ erg-sec from the experimental
blackbody data.\cite{Hermann} Planck's subsequent work (attempting to give a
statistical mechanical basis for his own interpolated fit to the experimental
data) separated the constant $\beta$ as $\beta=\hbar/k_{B}$ involving
Boltzmann's constant $k_{B}$ (related to the gas constant $R$ and Avogadro's
number $N_{A}$ of classical kinetic theory) and his constant $h=2\pi\hbar.$
\ Although $k_{B}=R/N_{A}$ involved familiar constants from classical kinetic
theory, the constant $\hbar$ was unfamiliar. \ Clearly, the constant $\hbar$
was a fundamental constant since it appeared in the universal blackbody
radiation spectrum. \ Indeed, Planck's constant $\hbar$ is an alternative
parameter to Stefan's constant $a_{S}$ which was introduced in 1879, since
Stefan's constant can be written in the form given in Eq. (\ref{Stef}).
\ However, Planck's constant $\hbar$ was a new fundamental constant in the
sense that it was unknown in classical statistical mechanics or in classical
electrodynamics. \ 

Indeed, in the struggle to understand blackbody radiation at the beginning of
the 20th century, one of the most compelling arguments for the turn to a new
quantum theory was the apparent absence of any place for the new Planck
constant $\hbar$ within classical mechanics or classical electrodynamics.
\ \ If Planck's constant is taken to zero, $\hbar\rightarrow0,$ then Planck's
spectrum in Eq. (\ref{Pl2}) becomes the Rayleigh-Jeans spectrum (\ref{equi})
to leading order in $\hbar\omega/(k_{B}T)$. \ Within classical physics, there
seemed to be no possible entry point for Planck's constant which was so vital
to Planck's blackbody radiation spectrum. \ It is noteworthy that Planck, the
\textquotedblleft reluctant revolutionary\textquotedblright\ of the early 20th
century, was desperately seeking a role for his constant $\hbar$ within
electrodynamics, and felt he could find none.\cite{Klein}

\subsection{Planck's Constant in Classical Electrodynamics}

Today, with the advantages of hindsight, we know exactly the classical
electrodynamic role for Planck's constant that eluded the physicists of the
early 20th century. \ Planck's constant appears naturally as the scale factor
in relativistic classical electromagnetic zero-point radiation. \ Yet it is
fascinating that our current textbooks of electrodynamics and modern physics
either omit mention of this natural role for $\hbar$ or hide the possibility
of a classical electrodynamic role for $\hbar.$

It was clear to the physicists of 1900, as it is clear to the physicists of
today, that Planck's constant $\hbar$ does not appear in Maxwell's equations,
the fundamental differential equations for the relativistic classical
electromagnetic fields. \ On the other hand, the speed of light $c$ is at
least implicit in Maxwell's equations. \ The general solution of Maxwell's
differential equations when written in terms of the electromagnetic potentials
$V(\mathbf{r},t)$ and $\mathbf{A(r},t)$ and using the retarded Green function,
takes the form
\begin{align}
V(\mathbf{r},t)  &  =V^{in}(\mathbf{r},t)+%
{\textstyle\int}
d^{3}r^{\prime}%
{\textstyle\int}
dt^{\prime}\frac{\delta(t-t^{\prime}-|\mathbf{r}-\mathbf{r}^{\prime}%
|/c)}{|\mathbf{r}-\mathbf{r}^{\prime}|}\rho(\mathbf{r}^{\prime},t^{\prime
})\nonumber\\
\mathbf{A}(\mathbf{r},t)  &  =\mathbf{A}^{in}(\mathbf{r},t)+%
{\textstyle\int}
d^{3}r^{\prime}%
{\textstyle\int}
dt^{\prime}\frac{\delta(t-t^{\prime}-|\mathbf{r}-\mathbf{r}^{\prime}%
|/c)}{|\mathbf{r}-\mathbf{r}^{\prime}|}\frac{\mathbf{J}(\mathbf{r}^{\prime
},t^{\prime})}{c} \label{Sols}%
\end{align}
where $\rho(\mathbf{r},t)$ and $\mathbf{J}(\mathbf{r},t)$ are the charge
density and current density of the sources, and $V^{in}(\mathbf{r},t)$ and
$\mathbf{A}^{in}(\mathbf{r},t)$ are the (homogeneous) source-free terms. \ All
the authors of the electromagnetism textbooks agree that these are indeed the
correct solutions. \ However, in current textbooks of electrodynamics, the
source-free terms $V^{in}(\mathbf{r},t)$ and $\mathbf{A}^{in}(\mathbf{r},t)$
are always omitted.\cite{omitted} \ Indeed, it is precisely as the scale
factor in these source-free terms that Planck's constant makes a natural entry
into relativistic classical electromagnetic theory.\cite{B2016c} \ \ Thus in a
shielded spacetime region at zero temperature where only source-free
contribution is zero-point radiation, the scalar potential can be taken to
vanish and the vector potential can be written as\cite{B1975a}%
\begin{align*}
\mathbf{A}(\mathbf{r},t)  &  =%
{\textstyle\sum_{\lambda=1}^{2}}
{\textstyle\int}
d^{3}k\widehat{\epsilon}(\mathbf{k},\lambda)\left(  \frac{\hbar}{2\pi
^{2}\omega}\right)  ^{1/2}\sin\left[  \mathbf{k}\cdot\mathbf{r-}\omega
t+\theta(\mathbf{k},\lambda)\right] \\
&  +%
{\textstyle\int}
d^{3}r^{\prime}%
{\textstyle\int}
dt^{\prime}\frac{\delta(t-t^{\prime}-|\mathbf{r}-\mathbf{r}^{\prime}%
|/c)}{|\mathbf{r}-\mathbf{r}^{\prime}|}\frac{\mathbf{J}(\mathbf{r}^{\prime
},t^{\prime})}{c}.
\end{align*}
The appearance of Planck's constant $\hbar$ in the source-free term
corresponds to the presence of relativistic classical electromagnetic
zero-point radiation.

From a historical perspective, we can see why it was so hard for the
physicists of the early 20th century to find a place for Planck's constant
within classical electrodynamics. \ Planck's constant does not enter the
fundamental differential equations for classical electrodynamics but rather
enters only in the (homogeneous) source-free solution to the differential
equations. \ Indeed, most current physicists of the early 21th century are
probably just as unaware of the possibility of a source-free contribution to
the general solution of Maxwell's equations as were the physicists of the
previous century. \ Our classical electromagnetism textbooks all omit the
(homogeneous) source-free contribution in their statement of the general
solution of Maxwell's equations, and so physicists are unaware of the
possibility. \ However, as has been pointed out before, the choice of this
(homogenous) source-free boundary condition is \textquotedblleft as much a
part of the postulates of the theory as the form of the Lagrangian or the
value of the electron charge.\textquotedblright\cite{Coleman}

\subsection{Contrasting Roles for Planck's Constant in Classical and Quantum
Theories}

Planck's constant enters both classical and quantum theories, but the roles
played in the theories are strikingly different. \ Within quantum theory,
Planck's constant is embedded in the foundations of the theory. \ The
existence of fundament commutators such as for position and momentum operators
$[\hat{x},\hat{p}_{x}]=i\hbar$ depends upon the non-zero value for $\hbar.$
\ From the commutators, one can derive the zero-point energy
$U_{\text{quantum}}$ for a quantum harmonic oscillator of natural frequency
$\omega_{0}$ giving $U_{\text{quantum}}=(1/2)\hbar\omega_{0}.$ \ Within
quantum theory, the quantum character of the theory and ideas of zero-point
energy both collapse with the limit $\hbar\rightarrow0$ which removes Planck's
constant from the theory. \ 

On the other hand, Planck's constant enters classical electrodynamics only as
a scale factor in the (homogeneous) source-free part of the general solution
of Maxwell's differential equations. \ Therefore there are two natural
versions of classical electrodynamics. \ In one form, Planck's constant
$\hbar$ is taken as non-zero, and zero-point radiation exists. \ In the other
form, the source-free contribution to the general solution of Maxwell's
equations is taken to vanish, and Planck's constant does not appear in the
theory. \ It is only this last version which is presented in the current
textbooks of classical electrodynamics and modern physics.

\section{Zero-Point Radiation, Nonrelativistic Physics, and the Rayleigh-Jeans
Consensus}

Unable to find a role for Planck's constant within classical mechanics or
classical electrodynamics, and confronted with classical derivations leading
to the Rayleigh-Jeans spectrum, the physicists of 1910 concluded that
classical physics led inevitably to the \textquotedblleft ultraviolet
catastrophe\textquotedblright\ contained within the Rayleigh-Jeans spectrum.
\ This consensus regarding blackbody radiation, which was reached a century
ago, is still repeated in all the textbooks.\cite{err}

\subsection{Collapse of the Arguments of 1910}

The thermal experiments of the early 20th century involved a crucial
limitation which confused the classical electromagnetic theorists at the
beginning of the 20th century. \ The experimentalists measured only the random
radiation of their sources which was above the random radiation surrounding
their measuring devices. If their sources were at the same temperature as
their measuring devices, the measuring devices registered no signal at all.
Today, in contrast to the end of the 19th century, random classical radiation
measurements are available which are of an entirely different character from
those of a century earlier.

Today, awareness of the possibility of relativistic classical electromagnetic
zero-point radiation (and indeed of its experimental detection by Casimir
force measurements), leads to the collapse of all the arguments of 1910 in
favor of the Rayleigh-Jeans spectrum being the inevitable conclusion of
classical physics. \ All the theoretical analysis for blackbody radiation in
the early 20th century was in terms of \textit{nonrelativistic} classical
statistical theory. \ However, nonrelativistic physics cannot support the idea
of a zero-point energy. \ Nonrelativistic classical statistical mechanics
cannot include zero-point radiation. \ Therefore it is not surprising that
application of classical statistical mechanics gave agreement with experiment
only in the long-wavelength (low frequency) region of the spectrum where the
relativistic zero-point contribution is negligible and can safely be ignored. \ 

\subsection{Nonrelativistic Scattering Calculations and the Importance of
Relativity}

There is a further set of calculations which did not influence the analysis of
the early 20th century but appeared subsequently and seemed to confirm the
earlier conclusion. \ These are the scattering experiments involving
nonrelativistic mechanical scatterers in random radiation. \ As was remarked
earlier, radiation equilibrium is achieved when some mechanical scatterer
(black particle) interacts with radiation in a cavity. \ A point dipole
oscillator does not act as a black particle since it only changes the
directions of the radiation and not the frequency spectrum.\cite{B1975a}
\ However, a charged \textit{nonlinear} oscillator is not passive, and will
indeed change the frequency spectrum of random radiation.

In 1925, van Vleck calculated the scattering of random radiation by a
nonrelativistic nonlinear multiply periodic system in the dipole approximation
which couples radiation to particle motion only through the frequency of the
motion, and not through a spatially-extended orbital motion. \ Although he
published only a preliminary report on his work,\cite{vanV} he concluded that
the nonlinear multiply periodic system assumed the Boltzmann distribution and
was in equilibrium with the Rayleigh-Jeans spectrum of random radiation.
\ This same conclusion was reached in subsequent calculations.\cite{nonlinosc}
\ Indeed, in 1983, Blanco, Pesquera, and Santos\cite{Blanco} went further and
showed that a charged particle with relativistic linear momentum
$\mathbf{p}=m\gamma\mathbf{v}$ in a general class of potentials also came to
equilibrium at the Boltzmann distribution for the particle in the presence of
the Rayleigh-Jeans spectrum for the radiation. \ Significantly, the general
class of potentials excluded the Coulomb potential, the only potential in
relativistic electrodynamics. \ 

All of these calculations fail to represent nature in the high-frequency
region of the spectrum because they are not relativistic calculations. \ None
of these calculations allow zero-point energy in the mechanical system.
\ Indeed, the one calculation which attempted to be relativistic excluded the
one potential, the Coulomb potential, which, when incorporated into
relativistic classical electrodynamics, is indeed relativistic. \ Crucially,
the Coulomb potential allows relativistic mechanical zero-point energy;
\textit{large} mass $m$ is associated with \textit{high} velocity and
\textit{high} frequency so that the mechanical zero-point energy associated
with large values of $mc^{2}/(k_{B}T)$ is matched consistently with high
frequency radiation where $\hbar\omega/(k_{B}T)$ is large and relativistic
zero-point radiation dominates the thermal radiation spectrum. \ Contrary to
this relativistic situation, in nonrelativistic potentials $V(\mathbf{r})$
where $V(0)$ is finite, a large mass is associated with low frequency oscillations.

\section{Classical Calculations Giving the Planck Spectrum with Zero-Point
Radiation}

In the section above, we noted that the classical theoretical analyses leading
to the Rayleigh-Jeans law are valid only in the low-frequency portion of the
spectrum; the calculations use nonrelativistic classical mechanics and/or do
not allow the presence of relativistic classical zero-point radiation.
\ However, there are several historical calculations which can be transformed
by the introduction of zero-point radiation in a valid context. \ Here we
review these historical calculations, and also mention additional analyses.
\ The calculations come from a wide variety of views, but all contribute to
the overwhelming evidence that relativistic classical electrodynamics
including relativistic classical zero-point radiation predicts the Planck
spectrum with zero-point radiation for blackbody radiation. \ 

\subsection{Einstein's Fluctuation Analysis for Radiation Modes}

\subsubsection{Original Einstein Calculation}

We consider first Einstein's radiation-fluctuation analysis\cite{Einstein1909}
of 1909. \ Einstein considered the fluctuations of radiation in a box, and,
connecting entropy with probability, arrived at the relation between the
entropy $S$ \ of a radiation normal mode and the average energy $U$ of the
mode%
\begin{equation}
\frac{\partial^{2}S}{\partial U^{2}}=\frac{-k_{B}}{\left\langle \varepsilon
^{2}\right\rangle } \label{d2S}%
\end{equation}
where $\left\langle \varepsilon^{2}\right\rangle $ corresponds to the mean
square fluctuation in the energy of the radiation normal mode. \ Now for
radiation, the mean-square energy fluctuation can be obtained from purely
classical wave theory\cite{Rice}\cite{fluct} as $\left\langle \varepsilon
^{2}\right\rangle =U^{2}$. \ Introducing this result into Eq. (\ref{d2S}), we
have
\begin{equation}
\frac{\partial^{2}S}{\partial U^{2}}=\frac{-k_{B}}{U^{2}}.
\end{equation}
If we integrate once with regard to $U$, we have%
\begin{equation}
\frac{\partial S}{\partial U}=\frac{k_{B}}{U}=\frac{1}{T}%
\end{equation}
which leads to exactly the Rayleigh-Jeans result $U=k_{B}T$. \ This was
Einstein's result.

\subsubsection{Modification to Include Zero-Point Radiation}

However, if relativistic classical electromagnetic zero-point radiation is
present, making the total energy $U=U_{T}+U_{zp},$ then the entropy $S$ should
be associated with only the thermal energy $U_{T}$ and not with the zero-point
energy $U_{zp}=(1/2)\hbar\omega$ in the normal mode. \ As noted above in
Section V C3, zero-point radiation involves vanishing entropy. \ Nevertheless,
both sources of energy contribute to the amplitude of the radiation field and
so to the fluctuations of the radiation associated with the normal mode.
\ Thus we should associate the entropy change with the fluctuations
$\left\langle \varepsilon_{T}^{2}\right\rangle $ above the zero-point
fluctuations\cite{B1969c}
\begin{equation}
\left\langle \varepsilon_{T}^{2}\right\rangle =U^{2}-U_{zp}^{2}=(U_{T}%
+U_{zp})^{2}-U_{zp}^{2}=U_{T}^{2}+2U_{T}U_{zp}. \label{fluct}%
\end{equation}
Therefore the connection of Eq. (\ref{d2S}) becomes%
\begin{equation}
\frac{\partial^{2}S}{\partial U^{2}}=\frac{-k_{B}}{U^{2}-U_{zp}^{2}}.
\end{equation}
Now integrating once with respect to $U,$ we have
\begin{equation}
\frac{1}{T}=\frac{\partial S}{\partial U}=\frac{k_{B}}{2U_{zp}}\ln\left(
\frac{U+U_{zp}}{U-U_{zp}}\right)  .
\end{equation}
Simplifying and taking the exponential so as to remove the logarithm while
inserting $U_{zp}=(1/2)\hbar\omega$, we find exactly the Planck spectrum with
zero-point radiation as in Eq. (\ref{Pl3}) above. \ Thus simply adding
relativistic classical electromagnetic zero-point radiation to Einstein's
analysis immediately gives us the full Planck spectrum with zero-point
radiation. \ Since Einstein (and his contemporaries) did not consider
classical electromagnetic zero-point radiation, he instead reinterpreted Eq.
(\ref{fluct}) in terms of photons of energy $2U_{zp}=2[(1/2)\hbar\omega
]=\hbar\omega$.

Only relativistic classical electromagnetic radiation enters this calculation;
there are no considerations involving relativistic particle behavior. \ The
inclusion of classical zero-point radiation leads directly to the Planck
spectrum with zero-point radiation.

\subsection{Moving Oscillator-Particle of Einstein and Hopf}

\subsubsection{Original Einstein-Hopf Calculation}

In 1910, Einstein continued his work on blackbody radiation beyond the
radiation-fluctuation analysis mentioned above. \ Wishing to drive home the
point that classical physics predicted the Rayleigh-Jeans spectrum for the
equilibrium between radiation and matter, Einstein and Hopf\cite{EH}
considered a particle of mass $m$ moving with one translational degree of
freedom and containing a small dipole oscillator of natural frequency
$\omega_{0}$ interacting with random classical radiation. The idea of the
calculation was to connect the translational motion of the oscillator to the
spectrum of random radiation which was providing impulses to the particle
through the interaction with the dipole oscillator. Einstein and Hopf required
that the average kinetic energy of the oscillator must be $(1/2)k_{B}%
T,$corresponding to the kinetic energy equipartition theorem. \ This
kinetic-energy equipartition for the translational motion of point particles
was regarded as the most secure aspect of classical statistical mechanics. \ 

The change $mv(t+\tau)-mv(t)$ in the translational momentum $mv(t)$ of the
particle during a short time interval $\tau$ was taken as due to a random
impulse $\Delta$ due to the random radiation interacting with the oscillator,
and a velocity-dependent damping $-Pv(t)\tau$ due to the retarding force on
the oscillator arising from the motion of the oscillator through the
doppler-shifted random radiation. \ Thus at the end of the short time interval
$\tau$, the momentum was%
\begin{equation}
mv(t+\tau)=mv(t)+\Delta-Pv(t)\tau\label{impul}%
\end{equation}
In equilibrium, the average translational energy of the particle should not
change in time, so that through first order in the time $\tau$
\begin{align}
\left\langle \lbrack mv(t+\tau)]^{2}\right\rangle  &  =\left\langle
[mv(t)+\Delta-Pv(t)\tau]^{2}\right\rangle \nonumber\\
&  =\left\langle [mv(t)]^{2}\right\rangle +\left\langle \Delta^{2}%
\right\rangle +2(m-P\tau)\left\langle v(t)\Delta\right\rangle -2mP\tau
\left\langle \lbrack v(t)]^{2}\right\rangle
\end{align}
Now the random impulse $\Delta$ is as often positive as negative so that
$\left\langle v(t)\Delta\right\rangle =0,$ and we find%
\[
0=\left\langle \Delta^{2}\right\rangle -2MP\tau\left\langle \lbrack
v(t)]^{2}\right\rangle .
\]
Einstein and Hopf calculated\cite{EH}\cite{B1969b} the square of the average
random impulse $\left\langle \Delta^{2}\right\rangle $ and also the retarding
force coefficient $P$ for a general spectrum of random radiation and found the
connection
\begin{equation}
\frac{4\Gamma\pi^{4}c^{4}\tau}{5\omega^{2}}\left[  \rho(\omega,T)\right]
^{2}-2\left\{  \frac{6c\pi^{2}\Gamma}{5}\left[  \rho(\omega,T)-\frac{\omega
}{3}\frac{\partial\rho(\omega,T)}{\partial\omega}\right]  \right\}  \tau
m\left\langle [v(t)]^{2}\right\rangle =0 \label{EH}%
\end{equation}
Now assuming that the equilibrium translational kinetic energy of the particle
was $(1/2)mv^{2}=(1/2)k_{B}T,$ they derived $\rho_{T}(\omega,T)=[\omega
^{2}/(\pi^{2}c^{3})]k_{B}T,$ the Rayleigh-Jeans spectrum. \ 

\subsubsection{Einstein-Stern Modification}

It is striking that Einstein was among the first to realize that the inclusion
of zero-point energy sharply changed the equilibrium between matter and
radiation. \ In the years 1910-1912, Planck developed his \textquotedblleft
second theory\textquotedblright\ of blackbody radiation which included a
zero-point energy $(1/2)\hbar\omega$ for oscillators in equilibrium with
radiation, but excluded any zero-point energy for the radiation field itself.
\ Planck and the other physicists of the time apparently made no connection
between the oscillator zero-point energy and relativity. \ Einstein and
Stern\cite{ES} picked up the idea of zero-point energy for an oscillator and
modified the earlier Einstein-Hopf calculation by including a zero-point
energy $\hbar\omega$ (no factor of $1/2$) for the oscillator. \ Thus instead
of $[\rho(\omega,T)]^{2}$ in $\left\langle \Delta^{2}\right\rangle ,$ they
introduced $\rho(\omega,T)\{\rho(\omega,T)+\hbar\omega^{3}/(\pi^{2}c^{3})\}$
where the last term correspond to the zero-point energy $\hbar\omega$ of the
oscillator. The differential equation (\ref{EH}) from the work of Einstein and
Hopf was modified to
\begin{equation}
\frac{4\Gamma\pi^{4}c^{4}\tau}{5\omega^{2}}\rho(\omega,T)\left[  \rho
(\omega,T)+\frac{\hbar\omega^{3}}{\pi^{2}c^{3}}\right]  -2\left\{  \frac
{6c\pi^{2}\Gamma}{5}\left[  \rho(\omega,T)-\frac{\omega}{3}\frac{\partial
\rho(\omega,T)}{\partial\omega}\right]  \right\}  \tau m\left\langle
[v(t)]^{2}\right\rangle =0. \label{ESM}%
\end{equation}
Now introducing $(1/2)mv^{2}=(1/2)k_{B}T,$ Einstein and Stern found that the
modified relation (\ref{ESM}) gave the Planck spectrum without any zero-point
energy for the radiation field, $\rho(\omega,T)=$ $[\omega^{2}/(\pi^{2}%
c^{3})]U_{P}(\omega,T)$ with $U_{P}(\omega,T)$ as given in Eq. (\ref{Pl2}).
\ The Einstein-Stern calculation attracted little notice as a classical
calculation since physicists had already decided that classical physics led to
the Rayleigh-Jeans spectrum, and zero-point energy for an oscillator was
regarded as part of the new physics of discrete quanta.\cite{Kuhn319} \ As far
as classical physics is concerned, this calculation (with zero-point energy
$\hbar\omega$ for the oscillator but not for the radiation) is actually
inconsistent; an oscillator comes to equilibrium with ambient random radiation
when the average energy of the oscillator matches that of the radiation modes
at the same frequency as the oscillator. \ 

Einstein and Stern also pointed out that the Planck spectrum for the
oscillator as given in Eq. (\ref{Pl2}) did not go over fully to
the\ Rayleigh-Jeans asymptotic form at high temperature, but rather involved%
\begin{equation}
U_{P}(\omega,T)=\frac{\hbar\omega}{\exp[\hbar\omega/(k_{B}T)]-1}=k_{B}%
T-\frac{1}{2}\hbar\omega+O\left(  \frac{\hbar\omega}{k_{B}T}\right)  .
\label{ESP}%
\end{equation}
The inclusion of a zero-point energy $(1/2)\hbar\omega$ for the oscillator
removed the temperature-independent term on the right-hand side of Eq.
(\ref{ESP}).

\subsubsection{Recent Modification}

Another modified version of the Einstein and Hopf calculation, but now
involving purely classical electromagnetic theory, appeared in
1969.\cite{B1969b} \ This time zero-point energy $(1/2)\hbar\omega$\ was
introduced in the classical radiation field with only the thermal part of the
radiation field regarded as unknown. \ Due to the zero-point radiation alone,
there is no velocity-dependent retarding force, since zero-point radiation is
relativistically invariant and takes the same form in every inertial frame.
\ Thus in the presence of zero-point radiation, the particle of Einstein and
Hopf (which involved no radiation damping of the translational motion) would
undergo a random walk in velocity without any retarding force, and would never
come to equilibrium. \ In this case, the Einstein-Hopf analysis required
modification associated with radiation damping of the translational motion as
the particle was accelerated at the walls of the container. \ Introducing the
radiation-damping impulse into Eq. (\ref{impul}) corresponds to removing the
impulse of the zero-point radiation; the Einstein Hopf equation is then
changed to
\begin{equation}
0=\frac{4\Gamma\pi^{4}c^{4}\tau}{5\omega^{2}}\left[  [\rho(\omega
,T)]^{2}-[\rho(\omega,0)]^{2}\right]  -2\left\{  \frac{6c\pi^{2}\Gamma}%
{5}\left[  \rho(\omega,T)-\frac{\omega}{3}\frac{\partial\rho(\omega
,T)}{\partial\omega}\right]  \right\}  \tau m\left\langle [v(t)]^{2}%
\right\rangle
\end{equation}
where $\rho(\omega,0)=[\omega^{2}/(\pi^{2}c^{3})](1/2)\hbar\omega$ corresponds
to the zero-point radiation spectrum. \ Now, the solution for the radiation
field giving the average kinetic energy $(1/2)mv^{2}=(1/2)k_{B}T$ for the
particle of large mass was found to be the Planck spectrum including
zero-point radiation $\rho(\omega,T)=$ $[\omega^{2}/(\pi^{2}c^{3}%
)]U_{Pzp}(\omega,T)$ with $U_{Pzp}(\omega,T)$ as given in Eq. (\ref{Pl3}).
\ Since this modified Einstein-Hopf calculation involves both relativistic
classical electromagnetic zero-point radiation and can be regarded as a
relativistic calculation involving negligible velocity in the large-mass
limit, the calculation meets all the criteria suggested here for giving an
experimentally-correct radiation spectrum, and indeed the correct spectrum is
what is found. \ 

It is also noteworthy that Marshall\cite{M1981} has provided a modified
version of Einstein's 1909 analysis\cite{Einstein1909} of the Brownian motion
of a moving mirror in the presence of random radiation. \ Again Einstein's
derivation of the Rayleigh-Jeans spectrum becomes a derivation of the Planck
spectrum with zero-point radiation when zero-point radiation is included in
the random thermal radiation.

\subsection{Comparing Diamagnetism of a Free Particle and Paramagnetic
Behavior}

According to the form of classical electrodynamics which ignores classical
zero-point radiation and which appears in all the textbooks, classical physics
does not allow diamagnetic behavior.\cite{Gdia} \ However, if classical
electromagnetic zero-point radiation is included, then classical
electromagnetism indeed shows diamagnetic behavior for a charged particle in
an isotropic harmonic-oscillator potential in the presence of a magnetic
field.\cite{MarRoyS} \ 

Although diamagnetic behavior for a charged particle does indeed appear within
classical physics including classical zero-point radiation, the diamagnetic
behavior disappears if the random radiation spectrum is taken as the
Rayleigh-Jeans spectrum.\cite{diam} \ Thus we again encounter distinct
asymptotic forms associated with low- and high-temperature limits. \ We expect
that in thermal radiation, the diamagnetic behavior is a continuous function
of temperature $T.$ \ The behavior is most striking for the free-particle case
involving a large magnetic field $B$ and a small isotropic oscillator
potential $V(r)=(1/2)m\omega_{0}r^{2},$ $eB/(mc)>>\omega_{0}.$ \ In this case
at zero temperature, the average particle orbital angular momentum
$\left\langle \mathbf{L}\right\rangle $ takes the magnitude $\hbar$ and is
oriented antiparallel to the direction of the magnetic field.\cite{diam} \ As
the mass $m$ is increased while maintaining the frequency ratio $eB/(mc)$ as
constant, the situation becomes that of negligible velocity, and so can be
regarded as the low-velocity limit of a relativistic system. \ 

Now the existence of a non-zero average magnetic moment at low temperature and
its disappearance at high temperature is exactly the sort of behavior which is
found for a paramagnetic magnetic moment treated using classical statistical
mechanics with a Boltzmann factor $\exp[-\mu B\cos\theta/(k_{b}T)]$ taken over
all possible orientations for the magnetic moment $\overrightarrow{\mu}.$ \ If
the paramagnetic moment is embedded in a spherical particle of very large
moment of inertia, then all frequencies should be very small, and hence
zero-point energy should not play any role; \ therefore applying
nonrelativistic classical statistical mechanics to the paramagnetic particle
should be justified. \ If we require that the ratio of the diamagnetic and
paramagnetic magnetic moments should be the same at all temperatures $T$, and
solve for the spectrum of random radiation which gives this behavior, then we
find\cite{diam} exactly the Planck spectrum with zero-point radiation in Eq.
(\ref{Pl3}).\ 

In this example yet again, we see the importance of including relativistic
classical electromagnetic zero-point radiation and also of being sure that the
behavior fits with relativistic electrodynamics. \ Again the Planck spectrum
with zero-point radiation appears naturally as the spectrum of radiation equilibrium.

\subsection{Fully Relativistic Analysis for Radiation in a Non-Inertial
Rindler Frame}

A further derivation of the Planck spectrum within classical physics with
classical zero-point radiation involves the use of ideas of conformal
transformations of free electromagnetic fields.\cite{Rindler} \ Classical
zero-point radiation corresponds to the spectrum of random classical radiation
which has the least possible information. \ Indeed, the correlation function
involving zero-point radiation depends upon only the geodesic separation
between the spacetime points at which it is evaluated. \ In an inertial frame,
Lorentz-invariant zero-point radiation is invariant under conformal
transformation. \ However, in a non-inertial frame, such as a uniformly
accelerating Rindler coordinate frame, a time-dilating conformal
transformation carries the zero-point radiation spectrum at $T=0$ into a
thermal spectrum at positive temperature $T>0$. \ It is then possible to go to
the asymptotic region of the coordinate frame where the spacetime becomes
Minkowskian, and so to recover Planck's spectrum including zero-point
radiation (\ref{Pl3}) as the blackbody radiation spectrum in an inertial
frame.\cite{Rindler} \ In this case, the entire analysis is fully relativistic
at every step.

\section{Closing Summary}

Blackbody radiation appeared in the physics research literature during the
first two decades of the 20th century, and accurate experimental measurements
of transferred heat energy made possible a comparison between theory and
experiment which seemed convincing. \ However, the physicists of the period
were unaware of two aspects which today are regarded as crucial to an
understanding of blackbody radiation within classical theory. \ One crucial
element is the presence of classical electromagnetic zero-point radiation with
its Lorentz-invariant spectrum. \ Significantly, the experimental work of that
early period did not measure the zero-point radiation, the theorists of the
period did not consider the possibility of classical zero-point radiation, and
accurate Casimir force measurements making clear the need for zero-point
radiation within classical theory did not occur until nearly a century later.
\ Furthermore, the scale of zero-point radiation is associated with Planck's
constant $\hbar$ which did not appear in 19th century mechanics or
electrodynamics. \ Indeed, the physicists of the early 20th century did not
see any way that Planck's constant could appear within a classical theory. \ 

Today we are aware that Planck's constant can appear naturally within
classical electrodynamics as the scale factor for the source-free part of the
general solution of Maxwell's equations. \ However, today most students see
Planck's constant as exclusively an element of quantum theory and are
completely unaware of its role as a scale for a Lorentz-invariant spectrum of
random classical radiation. \ In sharp contrast with quantum theory which
depends upon Planck's constant $\hbar$ for its basic algebraic structure,
classical electromagnetism makes use of Planck's constant only as the
scale-factor of the (homogeneous) source-free part of the general solution of
Maxwell's differential equations. \ Thus classical electrodynamic theory can
exist either with or without Planck's constant. \ In order to describe as many
aspects as possible in nature while using classical theory, we must include
classical zero-point radiation. \ Indeed, Planck's constant appears in
classical derivations of Casimir forces, van der Waals forces, low-temperature
specific heats of solids, diamagnetism, atomic structure, and blackbody
radiation.\cite{any}

The second crucial element of classical theory unrecognized by the physicists
of a century ago is the importance of relativistic behavior. \ Although it
became clear during the 19th century that Newtonian mechanics and classical
electrodynamics were in conflict, it was only at the turn of the 20th century
that classical electrodynamics was recognized as satisfying Lorentz invariance
whereas nonrelativistic mechanics satisfied Galilean invariance. \ However, in
the histories of the blackbody problem, there is no suggestion that physicists
ever considered any connection between relativity and the blackbody problem.
\ The same situation holds true today in the textbooks of modern physics where
special relativity is presented without making any connection to the problem
of blackbody radiation. \ Attempts to use nonrelativistic classical
statistical mechanics or the scattering of radiation by nonrelativistic
systems all led simply to the low-frequency Rayleigh-Jeans portion of the
blackbody spectrum because nonrelativistic classical mechanical systems cannot
allow the presence of relativistic zero-point radiation which is required for
an understanding of the full Planck blackbody spectrum within classical
physics. \ Although point harmonic oscillators can be fitted into relativistic
classical electrodynamics as mechanical systems involving negligible velocity,
nonlinear mechanical systems cannot be regarded as relativistic unless they
correspond to the Coulomb interaction which is part of relativistic classical
electrodynamics. \ Only the relativistic Coulomb potential with the
fundamental constant $e^{2}/c$ allows a relativistic mechanical zero-point
energy which fits with the relativistic zero-point radiation of classical
electromagnetic theory. \ 

Because of the failure to consider relativity and zero-point radiation, during
the first two decades of the 20th century, physicists came to an erroneous
conclusion regarding the classical physics of blackbody radiation, and this
erroneous conclusion is still repeated in the textbooks today. \ The
physicists of the early 20th century concluded that classical physics led
inevitably to the Rayleigh-Jeans spectrum for the full spectrum of thermal
radiation. \ This false conclusion arose because physicists did not (and still
do not) realize the important implications of classical zero-point radiation
and special relativity.

(revised November 3, 2017)


\begin{thebibliography}{99}
\bibitem {hbar}We will refer to both $\hbar$ and $h=2\pi\hbar$ as
\textquotedblleft Planck's constant.\textquotedblright\ \
\bibitem {quant}It is noteworthy that within quantum mechanics, there is no
thought of consistent $\sigma_{ltU^{-1}}$-scaling between mechanical systems
and zero-point radiation. Thus quantum mechanics involves only a
$\sigma_{tU^{-1}}$-scaling connecting time and energy, with a completely
independent scaling in length. \
\bibitem {Kuhn319}See ref. 19, p. 319, note 29. \ Kuhn mentions this
Einstein-Stern calculation only in the notes where he includes Einstein and
Stern's closing remark. \ At the end of their successful calculation of
Planck's spectrum, they comment, \textquotedblleft It seems doubtful that the
other difficulties will be overcome without the hypothesis of
quanta.\textquotedblright
\end{thebibliography}
\end{document}